\documentclass[11pt]{article}
\usepackage{geometry}                
\usepackage{graphicx}
\usepackage{rawfonts}
\usepackage{amssymb}
\usepackage{epstopdf}
\usepackage{amsmath,amsthm,amssymb}
\usepackage{mathtools}
\usepackage{fullpage}
\usepackage{color}
\usepackage[usenames,dvipsnames,svgnames]{xcolor}
\usepackage[colorlinks=true,citecolor=blue,linkcolor=red,urlcolor=blue]{hyperref}
\usepackage[font=footnotesize, labelfont=bf, margin=0.5cm]{caption}
\usepackage[labelformat=simple]{subcaption}
\usepackage{rawfonts}
\usepackage{etex}

\usepackage{ulem}

\input prepictex.tex
\input pictex.tex
\input postpictex.tex
\include{macros-pictex}

\newtheorem{theo}{Theorem}
\newtheorem{lemma}{Lemma}

\newtheorem{remark}{Remark}

\def\Pr{\noindent \textit{Proof: }}
\def\qed{$\Box$}

\usepackage{tikz}

\def\2;{\;\;}

\def\IntZ{{\mathbb Z}}

\title{Self-avoiding walks and polygons confined to a square}

\author
{S.~G.~Whittington\thanks{\href{mailto:swhittin@chem.utoronto.ca}{stuart.whittington@utoronto.ca}}\\
\small Department of Chemistry\\
\small University of Toronto, Toronto, Ontario M5S 3H6, Canada
}

\begin{document}

\maketitle

\begin{abstract}
We prove several rigorous results about the asymptotic behaviour of the numbers of polygons and self-avoiding walks confined to a square on the square lattice.  Specifically we prove that the dominant asymptotic behaviour of polygons confined to an $L \times L$ square is identical to that of self-avoiding walks that cross an $L \times L$ square from one corner vertex to the opposite corner vertex.  We prove results about the sub-dominant asymptotic behaviour of self-avoiding walks  crossing a square and polygons confined to a square and extend some results to self-avoiding walks and polygons in a hypercube in $\IntZ^d$.  
\end{abstract}

\maketitle

\section{Introduction}
\label{sec:Introduction}
There has been renewed interest in problems associated with self-avoiding walks on the square lattice confined to a square \cite{BradlyOwczarek,GuttmannJensenOwczarek}. The original version of the problem asked for the number of self-avoding walks confined to an $L \times L$ square and crossing the square from one corner vertex to the opposite corner vertex \cite{AbbottHanson,GuttmannWhittington}.  If the number of self-avoiding walks crossing the square from corner to opposite corner is $C(L)$ then it is known rigorously \cite{AbbottHanson,Madras,GuttmannWhittington} that
\begin{equation}
C(L) = \lambda^{L^2 + o(L^2)}.
\label{eqn:walksasymptotics}
\end{equation}
The value of $\lambda$ is known to quite high accuracy from exact enumeration and series analysis.  References can be found in \cite{GuttmannJensenOwczarek}.

Recently the related problem of self-avoiding walks confined to a square with no restrictions on the end points has been investigated \cite{BradlyOwczarek,GuttmannJensenOwczarek}.  If the number of these walks is $W(L)$ then it is known rigorously \cite{GuttmannJensenOwczarek} that

\begin{equation}
W(L)= \lambda^{L^2 + o(L^2)}, 
\label{eqn:walksunrestricted}
\end{equation}
 with the same value of $\lambda$, so that the two classes of walks have the same dominant asymptotic behaviour.  Guttmann \textit{et al} \cite{GuttmannJensenOwczarek} also studied the number of polygons confined to a square and they found numerical evidence that the number of polygons, $P(L)$, has the same dominant asymptotic behaviour.  In this paper, we prove this rigorously in Section \ref{sec:polygons}.

As part of their numerical analysis, Guttmann \textit{et al} \cite{GuttmannJensenOwczarek}
 \textit{assumed} that $C(L)$ behaves as
\begin{equation}
C(L) = \lambda^{L^2 + bL +O(\log L)}
\label{eqn:conjecture}
\end{equation}
and they found numerical evidence that $W(L)$ and $P(L)$ behave in the same way with the same value of $b$.  They estimated $b$ to be a small negative number.  Towards this, in Section \ref{sec:saws}, we prove that $C(L)$ behaves as
\begin{equation}
C(L) = \lambda^{L^2 + O(L)},
\end{equation}
and we extend this result to polygons in Section \ref{sec:polygonssub}.

Some of these results can be extended to confinement in a hypercube in the $d$-dimensional hypercubic lattice, $\IntZ^d$.  We explore this issue in Section \ref{sec:higherdimensions}.

\section{Polygons confined to a square}
\label{sec:polygons}

In this section we give a rigorous proof that the dominant asymptotics for polygons confined to a square is the same as for self-avoiding walks crossing a square.  We consider three cases:
\begin{enumerate}
\item
Polygons confined to an $L \times L$ square but with no other restrictions.  That is, the polygons are not required to have edges in the bounding sides of the square.  We call the number of these polygons in an $L \times L$ square $P(L)$.
\item
Polygons confined to the square and required to have at least one edge in the top and bottom sides of the square.  We call the number of these polygons $P_2(L)$.  Clearly the number of  polygons required to have at least one edge in each of the left and right sides of the square is also $P_2(L)$.
\item
Polygons confined to the square and required to have at least one edge in each of the four sides of the square.  We call the number of these polygons $P_4(L)$.
\end{enumerate}
By inclusion, $P_4(L) \le P_2(L) \le P(L)$.

The proof works by proving upper and lower bounds on $P(L)$.  We first note that $P(L)$ is a strictly monotone increasing function of $L$.  Clearly any polygon that fits in an $L \times L$ square will also fit in an $(L+1) \times (L+1)$ square.  But there exist polygons that fit in an $(L+1) \times (L+1)$ square (for instance polygons with an edge in each of the boundary sides of the square) but not in an $L \times L$ square.  Therefore $P(L-1) < P(L) < P(L+1)$.

We next give a lower bound.
\begin{lemma}
$P(L)$ satisfies the inequality
$$\liminf_{L\to\infty} \frac{1}{L^2} \log P(L).  \ge \liminf_{L\to\infty} \frac{1}{L^2} \log P_4(L)    \ge \log  \lambda.$$
\label{lem:lowerbound}
\end{lemma}
\Pr
To get a bound on $P(L)$ consider four $L \times L$ squares, arranged to form a $(2L+1) \times (2L+1)$ square with one lattice space between the adjacent pairs of squares.  Call these four squares NW, NE, SE and SW in an obvious notation.  Consider self-avoiding walks in each of the four squares, crossing the NW and SE squares from the bottom left to the top right vertex, and crossing the NE and SW squares from the top left to the bottom right  vertex.  By joining up four walks, one from each square with four additional edges we get a polygon in the $(2L+1) \times (2L+1)$ square so
\begin{equation}
P(2L+1) \ge P_4(2L+1) \ge C(L)^4.
\end{equation}
Take logarithms and divide by $(2L+1)^2$ giving
\begin{equation}
 \frac{1}{(2L+1)^2} \log P(2L+1) \ge           \frac{1}{(2L+1)^2} \log P_4(2L+1) \ge \frac{4}{(2L+1)^2} \log C(L).
\end{equation}

Let $L\to\infty$.  Then
\begin{equation}
\liminf_{L\to\infty} \frac{1}{L^2} \log P(L) \ge     \liminf_{L\to\infty} \frac{1}{L^2} \log P_4(L) \ge \lim_{n\to\infty} \frac{1}{L^2} \log C(L) = \log \lambda.
\end{equation}
\qed

The corresponding upper bound is given by the following lemma.
\begin{lemma}
\label{lem:upperbound}
$P(L)$ satisfies the inequality
$$\limsup_{L\to\infty} \frac{1}{L^2} \log P(L) \le  \log \lambda.$$
\end{lemma}
\Pr
Deleting an edge in each polygon in an $L \times L$ square gives a walk with one less edge in the same square, so
$P(L) \le W(L)$.  
Taking logarithms, dividing by $L^2$ and letting $L \to \infty$ gives
\begin{equation}
\limsup_{L\to\infty} \frac{1}{L^2} \log P(L) \le \lim_{n\to\infty} \frac{1}{L^2} \log W(L) = \log \lambda
\end{equation}
where in the last step we have used the recent result (\ref{eqn:walksunrestricted}) due to Guttmann \textit{et al} \cite{GuttmannJensenOwczarek}.
\qed

The two lemmas taken together prove that
\begin{equation}
\lim_{L\to\infty} \frac{1}{L^2} \log P_4(L) =    \lim_{L\to\infty} \frac{1}{L^2} \log P_2(L) =     \lim_{L\to\infty} \frac{1}{L^2} \log P(L) = \log \lambda,
\end{equation}
so that polygons in a square have the same dominant asymptotic behaviour as self-avoiding walks crossing a square.

\section{Self-avoiding walks crossing a square}
\label{sec:saws}

In this section we prove a result about the subdominant asymptotic behaviour of self-avoiding walks that cross a square.  We prove the following theorem:
\begin{theo}
For self-avoiding walks crossing an $L \times L$ square the asymptotic behaviour is given by
$$C(L) = \lambda^{L^2 + O(L)}.$$
\label{theo:sqsaws}
\end{theo}
\Pr
The proof works by a construction that is related to one used by Bousquet-M\'{e}lou \textit{et al} \cite{Bousquet-Melou} and is a modification of that used by Whittington and Guttmann \cite{GuttmannWhittington}.  Fix $L$ and consider $k^2$ squares of side $L$, with $k$ odd, arranged in a square so that adjacent squares are unit distance apart.  They form a larger square of side $kL + k-1 = k(L+1) -1$.  Label the $L \times L$ squares $(i,j)$, $i,j = 1, 2 \ldots k$.  If $i+j$ is even, the $(i,j)$ square is crossed by a self-avoiding walk from the bottom left to the top right vertex of the square.  If $i+j$ is odd, the $(i,j)$ square is crossed by a self-avoiding walk from the top left to the bottom right vertex of the square.

The walks in the $L \times L$ squares are concatenated by adding an edge between adjacent pairs of squares in the order
$$(1,1),(2,1), \ldots (k,1),(k,2),(k-1,2), \ldots (1,2),(1,3), (2,3), \ldots (k,3),(k,4), \ldots (k,k).$$
This procedure produces a walk in a $(k(L+1)-1) \times (k(L+1)-1)$ square, so that
\begin{equation}
C(kL+k-1) \ge C(L)^{k^2}.
\label{eqn:walksinequality}
\end{equation}
Take logarithms, divide by $(kL+k)^2$ and let $k\to\infty$ giving
\begin{equation}
\log \lambda \ge \frac{ \log C(L)}{(L+1)^2}
\end{equation}
or
\begin{equation}
C(L) \le \lambda^{L^2 + 2L +1}.
\end{equation}
The inequality $\lambda^{L^2} \le 2 C(L+3)$ follows from the results in \cite{Bousquet-Melou}, Section 4,  and implies that $C(L) \ge \frac{1}{2} \lambda^{(L-3)^2}$.  Hence $ \frac{1}{2} \lambda^{L^2 -6L+9} \le C(L) \le \lambda^{L^2 + 2L + 1}$, completing the proof.
\qed

\begin{remark}
The inequality in (\ref{eqn:walksinequality}) gives a simpler route to proving the asymptotic result (\ref{eqn:walksasymptotics}) than that given in \cite{GuttmannWhittington}, as we shall see in Section \ref{sec:higherdimensions}.
\end{remark}

\section{Sub-dominant asymptotics for polygons in a square}
\label{sec:polygonssub}

We next investigate the sub-dominant asymptotic behaviour of polygons in a square.  Consider an $L \times L$ square with the four corner vertices of the square at $(0,0)$, $(L,0)$, $(L,L)$ and $(0,L)$.  Define the \textit {top edge} of the polygon as the left-most edge in the top row of edges of the polygon.

\begin{theo}
For polygons confined to a square and with restrictions about having edges in the boundary of the square
$P_2(L) = \lambda^{L^2 + O(L)}$  and  $ P_4(L) = \lambda^{L^2 + O(L)}.$
\label{theo:polygonssub}
\end{theo}
\Pr
Recall that $P_4(L) \le P_2(L)$.  Consider polygons restricted to the square with at least one edge in each of the top and bottom boundaries of the square, counted by $P_2(L)$.  Delete the top edge of the polygon giving two vertices of degree 1.  Suppose the coordinates of these vertices are $(x,L)$ and $(x+1,L)$, with $0 \le x < L$.  Add an edge from $(x,L)$ to $(x,L+1)$ and a sequence of edges (if necessary)  from $(x,L+1)$ to $(0,L+1)$.  Add an edge from $(x+1,L)$ to $(x+1,L+1)$, followed by a sequence of edges from $(x+1,L+1)$ to $(L+1,L+1)$ and a sequence of edges from $(L+1,L+1)$ to $(L+1,0)$.  This construction results in a self-avoiding walk crossing the $(L+1) \times (L+1)$ square from the northwest corner vertex to the southeast corner vertex.  Each polygon gives a distinct walk so we have the inequality
\begin{equation}
P_4(L) \le P_2(L) \le C(L+1).
\end{equation}
Using the result from Theorem \ref{theo:sqsaws} we have 
\begin{equation}
P_4(L) \le P_2(L) \le \lambda^{(L+1)^2 + 2(L+1) + 1} = \lambda^{L^2 + 4L + 4}.
\label{eqn:polygonbound}
\end{equation}
To get a lower bound on $P_4(L)$ consider self-avoiding walks crossing an $(L-1)\times (L-1)$-square with opposite vertices at $(0,1)$ and $(L-1,L)$.  Add an edge from 
$(L-1,L)$ to $(L,L)$, a sequence of edges from $(L,L)$ to $(L,0)$, a sequence of edges from $(L,0)$ to $(0,0)$ and an edge from $(0,0)$ to $(0,1)$.  This gives a polygon in an $(L \times L)$-square with an edge in each side of the square so that $P_2(L) \ge P_4(L) \ge C(L-1) = \lambda^{(L-1)^2 + O(L-1) } = \lambda^{L^2 + O(L)}$.  
This result together with (\ref{eqn:polygonbound}) shows that
$P_2(L) = \lambda^{L^2 +O(L)}$ and $P_4(L) = \lambda^{L^2 +O(L)}$.
\qed

We now look at the case of polygons confined to a square but without restrictions on having edges in the boundary of the square.

\begin{theo}
\label{theo:snug}
For polygons confined to an $L \times L$ square
$$P(L) = \lambda^{L^2 + O(L)}.$$
\label{eqn:allpolygonbound}
\end{theo}
\Pr
If the polygon does not cross the square (\textit{ie} have at least one edge in each of two opposite sides of the square) then it will fit into a smaller \textit{snug} square that it crosses in at least one direction..  Suppose the snug square is $(L-k) \times (L-k)$.  The number of polygons crossing this snug square is bounded above by
$$2 \lambda^{(L-k)^2 + 4(L-k) + 4},$$
where we have used the result in (\ref{eqn:polygonbound}).
Each of these smaller squares can appear in the $L \times L$ square in less than $L^2$ ways.
Summing over $k$ gives
\begin{equation}
 P(L) \le \sum_{k=0}^{L-1}  2 L^2 \lambda^{(L-k)^2 + 4(L-k) + 4} \le 2L^3 \lambda^{L^2 + 4L +4} =  \lambda^{L^2 + O(L)}.
 \end{equation}
Since $P(L) \ge P_4(L) =  \lambda^{L^2 + O(L)}$ then $P(L) = \lambda^{L^2 + O(L)}$.
\qed

It is interesting to investigate the relationship between $P_2(L)$ and $P(L)$.  How similar are they for large $L$?  To do this we need to estimate the number of polygons that do not cross the square, \textit{ie} that have a span less than $L$ in both directions.
We know that $P(L) = \lambda^{L^2 + O(L)}$. If we can write $P(L) = \lambda^{L^2 + bL + o(L)}$ (a stronger statement) then we can deduce a useful result.  Polygons that do not cross the square must fit in a smaller square of side $L-1$,  although they do not have to cross this square.  Hence the number of polygons that do not cross the square is bounded above by
$$4 P(L-1)  =  \lambda^{(L-1)^2 + b(L-1) + o(L-1)}.$$
If we compare this with $P(L)$ we have 
$$\frac{4 P(L-1)}{P(L)} =  \lambda^{(L-1)^2 + b(L-1) -L^2 -bL +o(L)} =  \lambda^{-2L.  +o(L)}$$
Hence 
$$P(L) \ge P_2(L) \ge \frac{1}{2} P(L)\left(1-  \lambda^{-2L+o(L)}\right),$$
and this implies that the $b$ coefficient of $L$ (analogous to (\ref{eqn:conjecture})) for $P_2(L)$ is the same as that for $P(L)$.  This agrees with the numerical estimates by Guttmann \textit{et al} \cite{GuttmannJensenOwczarek}.


\section{Higher Dimensions}
\label{sec:higherdimensions}

Some of these results can be extended to higher dimensions.
Consider the hypercubic lattice $\IntZ^d$.  Concatenate $k^d$ hypercubes of side $L$ to form a larger hypercube of side $kL + k -1$ in an analogous way to the construction used in Section \ref{sec:saws}.  This gives the inequality
\begin{equation}
C(kL + k -1) \ge C(L)^{k^d}.
\label{eqn:ineqgend}
\end{equation}

\begin{theo}
For the hypercubic lattice $\IntZ^d$, if $C(L)$ is the number of walks that cross a hypercube of side length $L$ from one corner vertex to the diagonally opposite corner vertex, then the limit
$\lim_{L \to \infty} \frac{1}{L^d} \log C(L) \equiv \log  \lambda_d < \infty$
exists. 
\label{theo:gendasymp}
\end{theo}

\Pr
The number of walks that cross a hypercube of side $L$ is bounded above by $(2d)^{d L^d + o(L^d)}$ so $\limsup_{L\to\infty}  \frac{1}{L^d} \log C(L) \equiv \log \lambda_d < \infty$.
For $\epsilon > 0$ there are infinitely many values of $L$ such that 
\begin{equation}
\frac{1}{L^d} \log C(L) \ge  \left(1-\frac{\epsilon}{2} \right) \log \lambda_d .
\end{equation}
Choose one of these $L$ values such that $L^d/(L+1)^d \ge 1 - \epsilon /2$.
Then (\ref{eqn:ineqgend}) gives
\begin{equation}
\frac{\log C(kL + k - 1)}{k^d (L+1)^d} \ge  \frac{L^d}{(L+1)^d } \frac{\log C(L)}{L^d} \ge  \left(1-\epsilon  \right) \log \lambda_d . 
\end{equation}
Let $k \to \infty$ with $L$ fixed, giving
\begin{equation}
\liminf_{M\to\infty} \frac{1}{M^d} \log C(M) \ge (1-\epsilon )  \log \lambda_d .
\end{equation}
Letting  $\epsilon \to 0$ establishes the existence of the limit.
\qed

The next theorem is about the subdominant asymptotic behaviour.
\begin{theo}
For self-avoiding walks crossing an $L \times L \times ....  \times L$ hypercube in $\IntZ^d$, 
$$C(L) \le \lambda_d^{L^d + dL^{(d-1)}+o(L^{(d-1) })}.$$
\label{theo:hypercubicsaws}
\end{theo}

\Pr
Equation (\ref{eqn:ineqgend}) gives $C(kL + k -1) \ge C(L)^{k^d}$.
Take logarithms, divide by $(kL+k)^d$ and let $k\to\infty$ giving
\begin{equation}
\log \lambda_d \ge \frac{ \log C(L)}{(L+1)^d}
\end{equation}
or
\begin{equation}
C(L) \le \lambda_d^{L^d + dL^{(d-1)}+o(L^{(d-1) })}.
\label{eqn:walksboundhighdim}
\end{equation}
\qed 

Similar results can be proved for polygons confined to a $d$-dimensional hypercube.  For the $d$-dimensional  hypercubic lattice install a coordinate system $(x_1,x_2, \ldots, x_d)$.  
 Write $P_2(L)$ for the number of polygons in a hypercube of side $L$ with at least one edge in each of at least one pair of opposite faces, analogous to the definition for $d=2$.  We have the following Lemma.

\begin{lemma}
For polygons confined to and crossing a $d$-dimensional hypercube of side $L$, 
$$P_2(L) \le C(L+1) \quad \mbox{and} \quad C(L) \le P_2(L+1).$$
\end{lemma}
\Pr
Suppose that the hypercube of side $L$ has a pair of opposite vertices at $(0,0, \ldots, 0)$ and at $(L,L, \ldots, L)$.  Construct a hypercube of side $L+1$ with opposite vertices at
$(0,0, \ldots, 0)$ and at $(L+1,L+1, \ldots, L+1)$.  Consider a polygon in the $L$-hypercube  and suppose that $k$ is the smallest integer such that the polygon has at least one edge in each of the faces $x_k=0$ and $x_k=L$.  Delete a specified edge in the polygon in $x_k=L$, and add two edges to form a self-avoiding walk with its two  vertices of degree 1 in $x_k=L+1$.  Using only edges in the bounding hyperplanes of the $(L+1)$-hypercube, connect the two vertices of degree 1 to $(0,0, \ldots, L+1)$ and to $(L+1,L+1, \ldots, 0)$, forming a self-avoiding walk that crosses the $(L+1)$-hypercube.  Since different polygons give distinct walks, $P_2(L) \le C(L+1)$, proving the first part of the Lemma.
To prove the second part, consider an $L \times L \times \ldots L$ hypercube with opposite vertices at $(0,0, \ldots, 0)$ and $(L,L, \ldots, L)$ with a self-avoiding walk crossing the hypercube from one of these vertices to the other.  Extend this hypercube to a hypercube of size $L+1$ with opposite vertices at $(0,0, \ldots, 0, -1)$ and
$(L+1, L+1, L+1, \ldots, L)$.  Extend the walk by adding an edge between $(0,0, \ldots, 0)$ and  $(0,0, \ldots, -1)$, then edges joining $(0,0,0, \ldots, -1)$ to
$(L+1,0,0, \ldots, -1)$, then edges joining $(L+1,0,0, \ldots, -1)$ to $(L+1,L+1,0, \ldots, -1)$, ..., and eventually $(L+1,L+1, \ldots, L+1, -1)$ to $(L+1,L+1, \ldots, L+1, L)$, a vertex of the larger hypercube.  Finally, add edges between $(L+1,L+1, \ldots, L+1, L)$, and $(L,L,L, \ldots, L)$ to form a polygon.  This construction establishes that $C(L) \le P_2(L+1)$ and completes the proof.
\qed

This Lemma, together with Theorem \ref{theo:gendasymp} implies that 
\begin{equation}
\lim_{L\to\infty} \frac{1}{L^d} \log P_2(L) = \log \lambda_d.
\end{equation}

\begin{theo}
For polygons confined to and crossing a $d$-dimensional hypercube of side $L$, 
$$P_2(L) \le  \lambda_d^{L^d  + 2d L^{d-1} + o(L^{d-1})}.$$
\end{theo}
\Pr
Since $C(L) \le \lambda_d^{L^d + dL^{(d-1)}+o(L^{(d-1) })}$, see equation (\ref{eqn:walksboundhighdim}),
the inequality $P_2(L) \le C(L+1)$ implies that
$$P_2(L) \le \lambda_d^{(L+1)^d  + d(L+1)^{d-1} + o(L^{d-1})}  = \lambda_d^{L^d  + 2d L^{d-1} + o(L^{d-1})}.$$
 \qed

\section{Discussion}
\label{sec:discussion}
We have established rigorously some results recently found in numerical studies of self-avoiding walks and polygons confined to a square.  We have shown that walks and polygons have the same dominant asymptotic behaviour and we have proved results about the subdominant asymptotic behaviour for walks crossing a square from a corner vertex to the opposite corner vertex and for polygons confined to a square.  We have also investigated the problems of self-avoiding walks and polygons crossing a hypercube in $\IntZ^d$.

\section*{Acknowledgement  }
The author would like to thank Tony Guttmann, Aleks Owczarek and Buks van Rensburg for helpful correspondence.  Since completing this paper the author has learned that Guttmann and Jensen (arXiv:2211.14482) have proved the main result in Section 2 by a different argument.


\end{document}